# Exponential scaling of single-cell RNA-seq in the last decade

Valentine Svensson[1, 2, *], Roser Vento-Tormo[2], Sarah A Teichmann[1, 2]

## Author Affiliations

[1]European Molecular Biology Laboratory, European Bioinformatics Institute (EMBL-EBI), Hinxton, Cambridge, UK.

[2]Wellcome Trust Sanger Institute, Hinxton, Cambridge, UK.

[*]To whom correspondence should be addressed

## Abstract

The ability to measure the transcriptomes of single cells has only been feasible for a few years, and is becoming an extremely popular assay. While many types of analysis and questions can be answered using single cell RNA-sequencing, a central focus is the ability to survey the diversity of cell types within a sample. Unbiased and reproducible cataloging of distinct cell types requires large numbers of cells. Technological developments and protocol improvements have fuelled a consistent exponential increase in the numbers of cells studied in single cell RNA-seq analyses. In this perspective, we will highlight the key technological developments which have enabled this growth in data.

## Introduction

Many biological discoveries have been made possible by the development of new methods and technological progress. For example the improvements made to microscopes in the 17th century allowed Robert Hooke and Anton van Leeuwenhoek to describe the cell as the structural unit of life for the very first time (Gest, 2004). Since then, many studies have focused on cell characterisation, redefining the cell, not only as structural, but also as functional unit of life (Arendt et al., 2016).

Cells have traditionally been classified by their morphology or by expression of certain proteins in functionally distinct settings (Mosmann et al., 1986; Orkin, 2000; Poulin et al., 2016). Changes in cellular activity and identity are in many cases a reflection of distinct gene expression programs (Ivanov et al., 2007; Zhu, 2010). Proteomic technologies have the advantage of assaying the final functional product of the gene expression, but at the single cell level proteomics assays are constrained to a limited, pre-selected repertoire of molecules, precluding an unbiased, comprehensive analysis of cell phenotypes. The transcriptome (the abundance of all transcribed RNAs in a cell) provides an alternative to classify and characterise cells at the molecular level (Kolodziejczyk et al., 2015; Trapnell, 2015).

All single cell RNA-seq protocols share a common initial step, where transcribed RNA from cells can be converted to cDNA. The next step is an amplification, using molecular biological methods such as polymerase chain reaction (PCR) or in vitro transcription (IVT). The subsequent steps, culminating in sequencing allow the expression level of gene products to be quantified.

Eberwine *et al* measured the expression of a handful of individual genes from single cells for the first time in 1992 (Eberwine et al., 1992) using a sophisticated approach based on *in vivo* reverse transcription (RT), followed by amplification through *in vitro* transcription (IVT). Later, simpler PCR based methods emerged (Lambolez et al., 1992), and scaled up the number of cells and genes assayed over the years (Peixoto et al., 2004; Sheng et al., 1994). Eventually, untargeted single cell mRNA (or cDNA) amplification techniques were developed, which allowed researchers to perform transcriptome wide studies using microarrays (Esumi et al., 2008; Kurimoto et al., 2006, 2007; Tietjen et al., 2003). Building upon this, Tang *et al* adopted the technologies to be compatible with high throughput DNA sequencing, allowing completely unbiased transcriptome-wide investigation of the mRNA in a single cell (Tang et al., 2009).

Early single cell experiments were motivated by the prospect of an in depth analysis of gene expression in a few precious cells (Brouilette et al., 2012; Ramsköld et al., 2012; Tang et al., 2009, 2010, 2011). A shift in the field came when Guo *et al* demonstrated that distinct cell types could be identified without pre-sorting by performing RT-qPCR of 48 genes in parallel on more than 500 cells (Guo et al., 2010), demonstrating the utility of measuring larger numbers of cells. This inspired the Linnarsson group (Islam et al., 2011) to develop methods explicitly targeted at assaying many cells in parallel using unbiased RNA sequencing, with the long term goal of eventually cataloging all neuronal cell types (Shapiro et al., 2013; Zeisel et al., 2015).

In the past few years, many sensitive and accurate single-cell RNA-sequencing (scRNA-Seq) protocols have been introduced (Svensson et al., 2017a). Here, we focus on the increase in the number of cells profiled per study, a key feature in our ability to catalogue cell types. We review the exponential growth of scale in single cell transcriptomics experiments, from tens of cells up to hundreds of thousands single cells per study, in less than a decade, and discuss the technological advances which have enabled this (**Figure 1, Supplementary Table 1**).

Many different incremental improvements have contributed to increasing the scale, such as e.g. decrease in reagent volume and consumable costs (Kolodziejczyk et al., 2015; Picelli et al., 2013). In this perspective we focus on solutions for three key challenges: non-specific amplification of mRNA, automatic isolation of cells and the ability to process many cells in parallel.

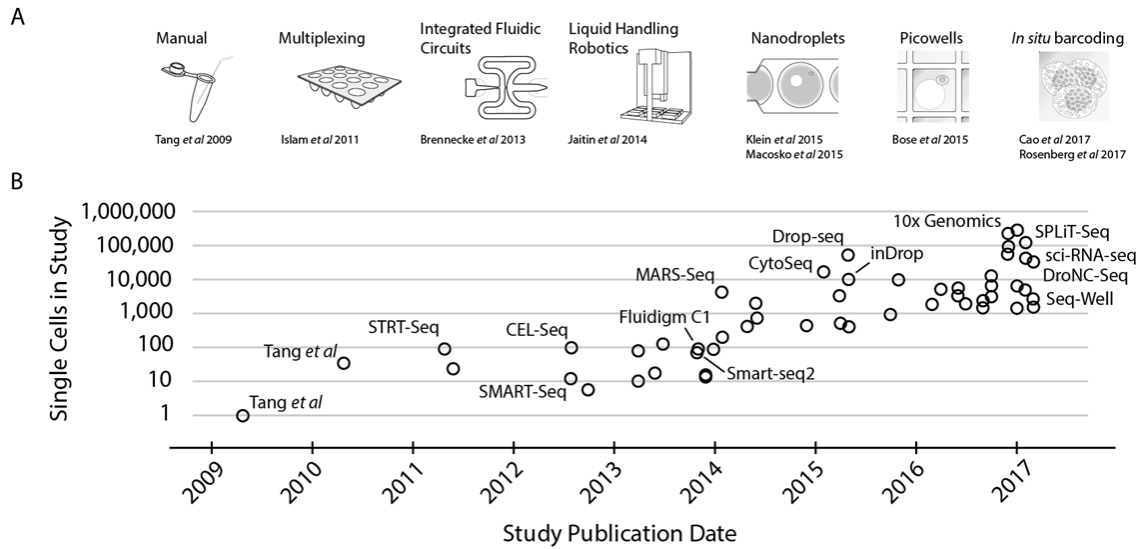

**Figure 1: Scaling of scRNA-seq experiments** (**A**) Key technologies allowing jumps in experimental scale. A jump to ~100 cells was enabled by sample multiplexing, a jump to ~1,000 cells by large scale studies using integrated fluidic circuits (IFCs), followed by a jump to several thousands using liquid handling robotics. Further order of magnitude jumps were enabled by random capture technologies through nanodroplets and picowell technologies. Recent studies have employed *in situ* barcoding to reach the next order of magnitude. (**B**) Cell numbers reported in representative publications by publication date. Key technologies and protocols are marked, and a full table with corresponding numbers is available in **Supplementary Table 1**.

## Whole transcriptome amplification from single cells

To successfully detect a signal when performing RNA sequencing, on the order of 0.1 - 1.0 µg of total RNA is needed[1,2]. The amount of RNA present in a single cell is limited, ranging from 1-50 pg depending on cell type (Boon et al., 2011). To overcome this problem, RNA can be converted into cDNA and amplified before creating a sequencing library. (DNA sequencing kits typically require on the order of 1 ng DNA[3]).

To achieve this, adaptor sequences need to be added to all mRNA transcripts. Delivery of these adaptors without pre-specifying target sequences of particular genes was the main technical challenge in early single cell RNA studies. Amplification of the cDNA can be exponential, by PCR amplification of the cDNA, or linear, via multiple rounds of IVT.

PCR amplification strategies require the addition of adaptor sequences at both ends of the double stranded cDNA. Most techniques utilise the polyA tail of the mRNA to generate the first strand of cDNA, by initiating reverse transcription using a poly(T)-oligonucleotide that also contains a universal adaptor sequence. The second adaptor is incorporated in the cDNA amplification step using different strategies, outlined below.

---

[1] Truseq, https://support.illumina.com/sequencing/sequencing_kits/truseq_rna_sample_prep_kit/input_req.html
[2] NEBNext, https://www.neb.com/faqs/2012/11/19/what-is-the-starting-material-i-need-to-use-when-preparing-libraries-using-the-nebnext-ultra-dire
[3] Nextera XT, https://support.illumina.com/sequencing/sequencing_kits/nextera_xt_dna_kit/input_req.html

In the homopolymer tailing approach, a transferase adds a poly(A) tail to the 3' end of the first-strand cDNA. Subsequently, a poly(T) primer with a different universal anchor is incorporated into the second strand. The two different adaptor sequences are then used for PCR amplification. This protocol was first applied to single-cell analysis in 2002 (Klein et al., 2002) and eventually adapted for use with microarrays (Kurimoto et al., 2006, 2007) and scRNAseq (Tang et al., 2009). Although long cDNAs have been transcribed by this approach (Tang et al., 2009), a drawback of the method is the reduction of 5' transcript coverage due to premature termination of the RT.

Template switching PCR is an alternative strategy that ensures the full transcription of the RNA. This method is based on the intrinsic ability of the transcriptases of the Moloney murine leukaemia virus (MMLV) to add a small number of nucleotides (mostly cytosines) when the RT reaches the end of the mRNA (Zhu et al., 2001). The addition of a helper oligonucleotide containing complementary nucleotides and the second adaptor, allow the polymerase to automatically initiate second strand synthesis without requiring the homopolymer tailing. In the single-cell tagged reverse transcription sequencing (STRT-Seq) method (Islam et al., 2011 Genome Research), full-length cDNA is amplified by template switching, but only the 5' end fragment is captured and sequenced. By contrast, fragments from the full-length cDNA is sequenced in the SMART-seq method (Ramsköld et al., 2012).

An alternative approach which was originally used in pioneering single-cell analysis of the transcriptome (Eberwine et al., 1992) is *in vitro* transcription (IVT). The CEL-Seq (Cell Expression by Linear amplification and Sequencing) protocol leveraged this for linear mRNA amplification from single cells. Here the RT adaptor also contains a T7 promoter, allowing the final double stranded DNA (dsDNA) to be transcribed into multiple copies of antisense RNA (aRNA) (Baugh et al., 2001). Once enough amplified aRNA has been produced it can be fragmented and again reverse transcribed to cDNA. This method was also used for MARS-seq (Jaitin et al., 2014) and later also for Drop-seq (Macosko et al., 2015) and Seq-Well (Gierahn et al., 2017).

# Simplified cell isolation

The earliest single cell experiments analysed only a handful of cells that were each manually isolated in single tubes. Concurrent with improvements in the methodology, larger numbers of cells were analysed in parallel, and microtiter plates were used for the new methods. Although some devices like the custom built semi-automatic cell picker (Islam et al., 2011) were designed, an easy to implement alternative was to use fluorescence assisted cell sorting (FACS) to isolate cells into microwell plates. FACS sorting has frequently been employed to increase the scale of single cell sequencing.

Some methods have made specialized adaptations of FACS sorting: the MASC-seq method makes use of a spatially barcoded Poly(dT) array which cells are sorted onto *(Vickovic et al., 2016)*. The recently published nanoliter-volume microwell array platform used for the STRT-seq-2i protocol also takes advantage of FACS to load cells into miniscule wells in custom capture plates.

A further increase in scale was achieved through robotic automation following cell capture (Jaitin et al., 2014; Muraro et al., 2016). However, specialised flow cytometers are expensive and require dedicated staff. Different forms of passive and random cell capture technologies were developed to reduce the need for such instrumentation.

The first form of passive cell capture for scRNA-seq was the microfluidic C1 system released by Fluidigm Inc. The company had expertise in producing systems for parallel RT-qPCR[4] in single cells, and adapted the technology for sequencing. The key feature of the Fluidigm technology is the design of microfluidics devices (or chips) that allow the sequential delivery of very small and precise volumes into tiny reaction chambers. Cells are loaded onto this chip and are passively captured in (up to) 96 isolated chambers in about half an hour. Several steps of the SMART-seq protocol could be performed within the chambers of the chip, before cDNA was extracted from the chip and deposited in microwell plates for the generation of a sequencing library. Complex single use IFC chips are fundamentally limited to a set number of chambers to capture cells in (96 in the current version), and in many cases only a fraction of the chambers successfully capture cells, depending on the cell type (Lönnberg et al., 2017; Zeisel et al., 2015). Some large scale studies made use of a large number of IFC's to create big data sets (Shalek et al., 2013; Zeisel et al., 2015).

More recently, methods emerged to randomly capture and manipulate individual cells in nanoliter droplet emulsions (Mazutis et al., 2013). The inDrop and Drop-seq protocols both present strategies to isolate cells in droplets and perform barcoded cDNA preparation within each droplet (Klein et al., 2015; Macosko et al., 2015). (Barcoding strategies are discussed below) Here, two flows of liquid, one containing reagents and beads with poly(T) RT primers, and another with cells in buffer, are merged into a combined flow. This flow is separated into droplets by addition of oil at intervals. Calibrating the relative rate of the two flows to each other, and to the creation of droplets, allows a user to ensure that in most cases only single cells will be isolated in droplets by Poisson statistics. The random nature of this process requires a large number of cells and as such is not suited to samples with limited availability of cells.

In an alternative strategy, random cells are deposited into picoliter wells that contain barcoded beads and reagents (Bose et al., 2015; Fan et al., 2015; Gierahn et al., 2017). The same beads as for droplet protocols can be used to deliver RT primers into individual wells, but no pumps or other microfluidic equipment are needed. Instead, cells are isolated into the picoliter wells by gravity. Each plate of wells will have an upper limit of cells determined by the number of wells on the plate, and Poisson statistics for the limiting dilution to avoid doublets, so that only a small fraction of the wells will end up containing cells.

# Barcoding strategies for efficient multiplexing

*Multiplexing* refers to the practice of adding molecular barcodes to cDNA fragments. This way material from many cells can be pooled together, allowing subsequent steps to be performed with a single tube. By performing the multiplexing step early in the workflow, the labour and material needed to create a sequencing library is dramatically reduced. It also enables the use of technologies which require a minimal input level of material, such as IVT (Hashimshony et al., 2012). Furthermore, libraries from many cells can be sequenced together to spread the sequencing read depth over many cells.

---

[4] https://www.fluidigm.com/about/aboutfluidigm

Islam *et al* published the first single-cell RNAseq protocol multiplexing cells from a single 96 well plate by using a unique template switching oligo (TSO) in each well in their STRT-seq method (Islam et al., 2011)[5]. In early multiplexing strategies, a unique barcode for each isolated cell is incorporated into the cDNA fragments before library generation; either in the TSO (Islam et al., 2011), the RT adapter (Hashimshony et al., 2012; Macosko et al., 2015; Zheng et al., 2017), or during full length PCR (Hochgerner et al., 2017).

Furthermore, different multiplexing steps can be combined. MARS-Seq, a variant of the CEL-seq protocol, adds an additional barcode during library preparation, to increase the number of cells that can be be sequenced in parallel. A similar strategy was introduced in the STRT-Seq-2i protocol (Hochgerner et al., 2017). Here up to 100 barcodes are first added during full-length PCR, followed by 96 additional barcodes for different regions of their array during library preparation, allowing a total of 9600 cells to be multiplexed and sequenced together.

Early multiplexing is critical for random cell isolation methods. To label cDNA from cells isolated in individual droplets (or picoliter wells), the beads with RT primers also contain a barcode. Reverse transcription and barcoding can therefore happen in the individual isolates. Following this, material can be treated as in other protocols. Drop-seq and Seq-Well uses the same IVT protocol as CEL-seq (Gierahn et al., 2017; Macosko et al., 2015) on the multiplexed material, while others use PCR based methods.

Allowing more cells to be multiplexed requires more barcodes, which in turn would require longer oligos at prohibitive synthesis cost. This can be avoided by two different strategies: combining multiple shorter barcodes into longer barcodes, or by synthesis of very long random barcodes.

The first approach, combinatorial combination of barcodes, was first described with the targeted CytoSeq method. The authors used initial pools of 96 barcodes into which beads were split-pooled 3 times to generate a set of $96^3$ = 884,736 barcoded beads (Fan et al., Science 2015). A similar method called "mix & expand" was later published by where the authors created $96^2$=9,216 barcoded beads (Bose et al., Genome Biology 2015). Similarly the InDrop method barcoded hydrogel beads by combinatorially ligating two sets of 384 barcodes, creating $384^2$ = 147,456 unique barcoded beads (Klein et al., 2015). This approach is also taken by the commercial Illumina SureCell system using $96^3$ = 884,736 unique beads[6].

The second approach was taken in the Drop=seq and Seq-well protocols, where the authors used 12 rounds of split-pool single base DNA synthesis on the beads to generate $4^{12}$ = 16,7 million randomer barcodes (Gierahn et al., 2017; Macosko et al., 2015). This procedure is simpler than the first approach, and do not require any synthesized oligos for the barcodes. However, in the first approach barcodes can be designed to avoid biases (Costea et al., 2013), and ensure barcode sequences will be distinct. To avoid potential barcode collisions randomer barcodes need to be longer, decreasing the probability that similar barcodes both get assigned to droplets with cells in them.

---

[5] A later version of STRT-seq removed early multiplexing in favour passive cell capture using the Fluidigm C1 platform.
[6] https://support.illumina.com/sequencing/sequencing_kits/surecell-wta-3-kit.html

# Current limitations and future directions

The isolation and handling of individual cells has become much simpler in recent years. A cost of this is the requirement of large cell suspensions. In labour intensive manual methods it can be ensured that as much cellular material as possible is used. When cells are randomly isolated in droplets or wells, only a fraction of the cells in the suspension will be captured.

We have not mentioned the challenge of creating cell suspensions. For many cell types this may not be an issue, but for certain tissues it can be difficult to make single cell suspensions, as different tissues will require different procedures. Additionally, storage of material could impose a challenge. Recent work has proposed moving to sequencing of individual nuclei to simplify creating suspensions, with positive results.

Cost of sequencing is still prohibitive, even at shallow depths, when studying hundreds of thousands of cells. Recent developments promise marginally cheaper sequencing from higher throughput[7], but some radical change in sequencing technology might be needed to bring down costs further.

The random isolation of cells we have described comes with inherent limitations. Poisson statistics of cell capture to ensure that mostly single cells are isolated means there will always be large inefficiencies in terms of cell isolation, and the pool of barcodes will always have to be substantially larger than the number of cells captured to avoid barcode duplication.

A remarkable recent strategy to avoid both of these pitfalls is combinatorial *in situ* barcoding. This technique was initially devised for single cell ATAC sequencing (Cusanovich et al., 2015) and later whole genome sequencing (Vitak et al., 2017) and Hi-C (Ramani et al., 2017). Recently, these concepts were adapted to scRNA-seq in the sci-RNA-seq (Cao et al., 2017) and SPLiT-Seq (Rosenberg et al., 2017) methods.

Here single cells are never individually isolated; instead cells are fixed and the mRNA is manipulated *in situ* inside each cell. Cells are split into mini-pools and distributed into multi-well plates with unique barcodes in each well. First strand synthesis labels all cells in the well with a first barcode. Cells are then pooled and again randomly split into mini-pools in plate wells, and a second well-specific barcode is then added. This procedure can be repeated *ad infinitum* before finally pooling the cells to amplify the material and create a sequencing library. This results in an arbitrarily low probability that any two cells will co-locate in the same sequence of wells, and so RNA from each cell is uniquely labelled.

Unlike isolation methods, the number of potential labelled cells in an experiment can be exponentially scaled by the number of barcoding rounds. This technology was also reported to be compatible with single nucleus sequencing (Rosenberg et al., 2017). Developments such as this hold the promise of enabling very large studies in the future, such as the catalogue of *C. elegans* cells in (Cao et al., 2017).

---

[7] https://www.illumina.com/systems/sequencing-platforms/novaseq.html

An important factor in cellular heterogeneity of tissues is the spatial context of cells (Lee, 2017). When creating a suspension of cells information about this is lost. Recently a method based on CEL-seq was created to retain the spatial location of mRNA expression in thin tissue slices (Ståhl et al., 2016), however not with single cell resolution. Other methods based on single molecule fluorescent *in situ* hybridisation (smFISH) have been scaled up to measure expression of up to 1,000 pre-defined target genes in tens of thousands of single cells using imaging (Moffitt et al., 2016a, 2016b; Shah et al., 2016) . Recent promising work has shown untargeted sequencing of cDNA from mRNA inside cells using the fluorescent *in situ* sequencing method (FISSEQ) (Lee et al., 2014, 2015). The ability to investigate the spatial context of gene expression will be important to understand cellular heterogeneity and enable us to identify genes important for tissue structure (Svensson et al., 2017b) and form a complementary technology to highly multiplexed sequencing of cell suspensions.

# Author Contributions

V.S., R.V-T., and S.A.T. wrote the manuscript. V.S. prepared the figure.

# Competing Financial Interests

The authors declare no competing financial interests.

# Acknowledgements

The authors would like to thank Zheng-Shan Chong, Kedar Nath Natarajan, Tomás Gomes, Natalie Edner, and Kerstin Meyer for reading the manuscript and giving helpful feedback.

R.V-T is supported by an EMBO Long-Term Fellowship and a Human Frontier Science Program Long-Term Fellowship.

# Figure legends

## Figure 1:

**Scaling of scRNA-seq experiments** (**A**) Key technologies allowing jumps in experimental scale. A jump to ~100 cells was enabled by sample multiplexing, a jump to ~1,000 cells by large scale studies using integrated fluidic circuits (IFCs), followed by a jump to several thousands using liquid handling robotics. Further order of magnitude jumps were enabled by random capture technologies through nanodroplets and picowell technologies. Recent studies have employed in situ barcoding to reach the next order of magnitude. (**B**) Cell numbers reported in representative publications by publication date. Key technologies and protocols are marked, and a full table with corresponding numbers is available in **Supplementary Table 1**.

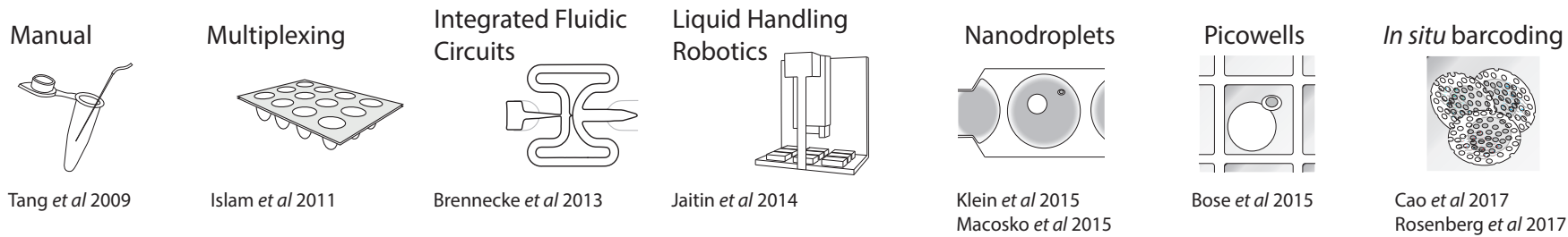
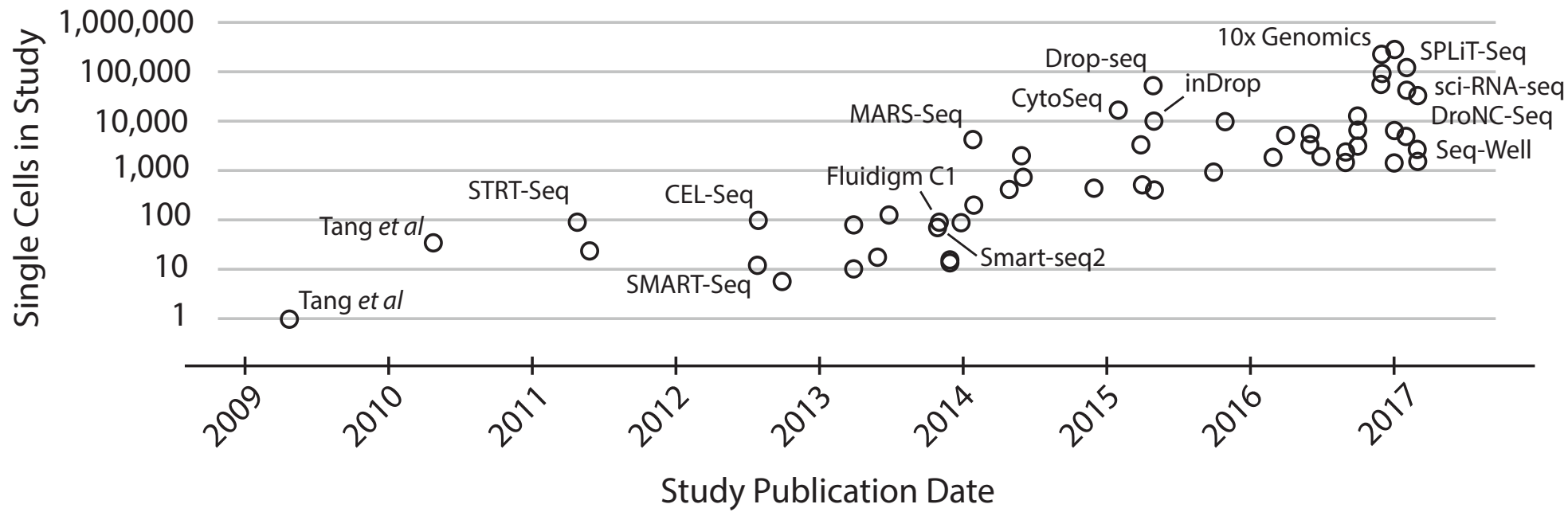